\documentclass[floats,floatfix,showpacs,amssymb,prd,twocolumn,superscriptaddress,nofootinbib,nolongbibliography,reprint]{revtex4-2}

\usepackage{amssymb,amsmath,verbatim,mathtools,needspace,enumitem,etoolbox,graphicx,physics,microtype,afterpage,bigints,gensymb,tabularx, comment}

\usepackage[dvipsnames, usenames]{xcolor}
\definecolor{linkcolor}{rgb}{0.0,0.3,0.5}
\definecolor{dodgerblue}{HTML}{1E90FF}
\usepackage[unicode, colorlinks=true, linkcolor=linkcolor, citecolor=linkcolor, filecolor=linkcolor,urlcolor=linkcolor, pdfusetitle]{hyperref}
\usepackage[all]{hypcap}
\usepackage[T1]{fontenc}
\usepackage[utf8]{inputenc}
\usepackage{orcidlink}
\usepackage[normalem]{ulem}

\newcommand{\ssim}{\mathchar"5218\relax\,}

\makeatletter
\newcommand*{\balancecolsandclearpage}{\close@column@grid \cleardoublepage \twocolumngrid}
\makeatother

\def\apj{{Astrophys.~J.}}\def\apjl{{Astrophys.~J.~Lett.}}
\def\apjs{{Astrophys.~J.~Supp.}}

\def\aap{{Astron.~Astrophys.}}

\def\mnras{{Mon.~Not.~R.~Astron.~Soc.}}

\def\prd{{Phys.~Rev.~D}}

\def\prx{{Phys.~Rev.~X}}
\def\prl{{Phys.~Rev.~Lett.}}

\def\natastro{{Nat. Astron.}}

\newcommand{\bham}{\affiliation{School of Physics and Astronomy \& Institute for Gravitational Wave Astronomy, University of Birmingham, \\ Birmingham, B15 2TT, United Kingdom}}
\newcommand{\milan}{\affiliation{Dipartimento di Fisica ``G. Occhialini'', Universit\'a degli Studi di Milano-Bicocca, Piazza della Scienza 3, 20126 Milano, Italy}}
\newcommand{\infn}{\affiliation{INFN, Sezione di Milano-Bicocca, Piazza della Scienza 3, 20126 Milano, Italy}}
\newcommand{\jhu}{\affiliation{William H. Miller III Department of Physics and Astronomy, Johns Hopkins University,
3400 North Charles Street, Baltimore, Maryland 21218, USA}}

\begin{document}

\title{Black-hole mergers in disk-like environments could\\explain  the observed $q - \chi_\mathrm{eff}$ correlation %
}

\author{Alessandro Santini$\,$\orcidlink{0000-0001-6936-8581}}
\email{a.santini6@campus.unimib.it}
\milan

\author{Davide Gerosa$\,$\orcidlink{0000-0002-0933-3579}}
\milan \infn \bham

\author{Roberto Cotesta$\,$\orcidlink{0000-0001-6568-6814}$\,$}
\jhu

\author{Emanuele Berti$\,$\orcidlink{0000-0003-0751-5130}$\,$}
\jhu


\date{\today}

\begin{abstract}
 Current gravitational-wave data from stellar-mass black-hole binary mergers suggest a correlation between the
  binary mass ratio $q$ and the effective spin $\chi_\mathrm{eff}$: more unequal-mass binaries consistently show larger and positive values of the effective spin. Multiple generations of black-hole mergers in dense astrophysical environments may provide a way to form unequal-mass systems, but they cannot explain the observed correlation on their own. We show that the symmetry of the astrophysical environment is a crucial feature to shed light on  this otherwise puzzling piece of observational evidence. We present a toy model that reproduces, at least qualitatively, the observed correlation.
  The model  relies on axisymmetric, disk-like environments where binaries participating in hierarchical mergers share a preferential direction. %
  Migration traps in AGN disks are a prime candidate for this setup, hinting at the exciting possibility of constraining their occurrence with gravitational-wave~data.
\end{abstract}

\maketitle

\section{Symmetry and black-hole mergers}
\label{sec:introduction}
The growing catalog of gravitational-wave (GW) observations by LIGO and Virgo~\cite{2019PhRvX...9c1040A, 2021PhRvX..11b1053A, 2021arXiv210801045T, 2021arXiv211103606T} provides a unique opportunity to understand and interpret the astrophysics of stellar-mass black-hole (BH) binaries. One of the most surprising features that emerged from recent data is a
correlation between the masses $m_i$ and dimensionless spins $\chi_i$ of the merging BHs, 
specifically between the binary mass ratio $q$ and effective spin $\chi_{\rm eff}$. These are defined  as
\begin{equation}
\label{definitions}
  q = \frac{m_2}{m_1}\leq 1\,, \qquad \chi_{\rm eff} = \frac{\chi_1 \cos{\theta_1} + q \chi_2 \cos{\theta_2}}{1 + q} \in [-1,1]\,,
\end{equation}
where $\theta_i$ is the angle between each spin vector and the orbital angular momentum of the binary $\boldsymbol{L}$~\cite{2001PhRvD..64l4013D}.
Binaries with small values of $q$ tend to have large and positive values of $\chi_{\rm eff}$.
This trend was first found
by \citeauthor{2021ApJ...922L...5C}~\cite{2021ApJ...922L...5C} %
and later confirmed with both a larger dataset~\cite{2023PhRvX..13a1048A} and a different statistical method~\cite{2022MNRAS.517.3928A,2023arXiv230715278A}. 

From an astrophysical standpoint, the observed correlation between mass ratio and effective spin is puzzling.
Unequal-mass binaries can naturally form if hierarchical mergers occur in dense astrophysical environments
(see Ref.~\cite{2021NatAs...5..749G} for a review). The remnants of BH mergers are more massive than their progenitors and, when paired with other BHs from the same stellar population, are  natural candidates to form binaries with mass ratio $q$
significantly smaller than unity. Moreover, the ``higher-generation'' BHs formed as a result of a previous merger are expected to show a characteristic spin distribution peaked at $\chi\ssim0.7$~\cite{2017PhRvD..95l4046G,2017ApJ...851L..25F}, which might translate into higher values of the effective spin $\chi_{\rm eff}$.
While hierarchical mergers could naturally pair  low mass ratios to large spin magnitudes, this does not explain why {\em only positive} values of the effective spin would preferentially be associated with unequal masses, as currently observed~\cite{2021ApJ...922L...5C,2023PhRvX..13a1048A}.

\begin{figure*}
		\centering
		\includegraphics[width=2\columnwidth]{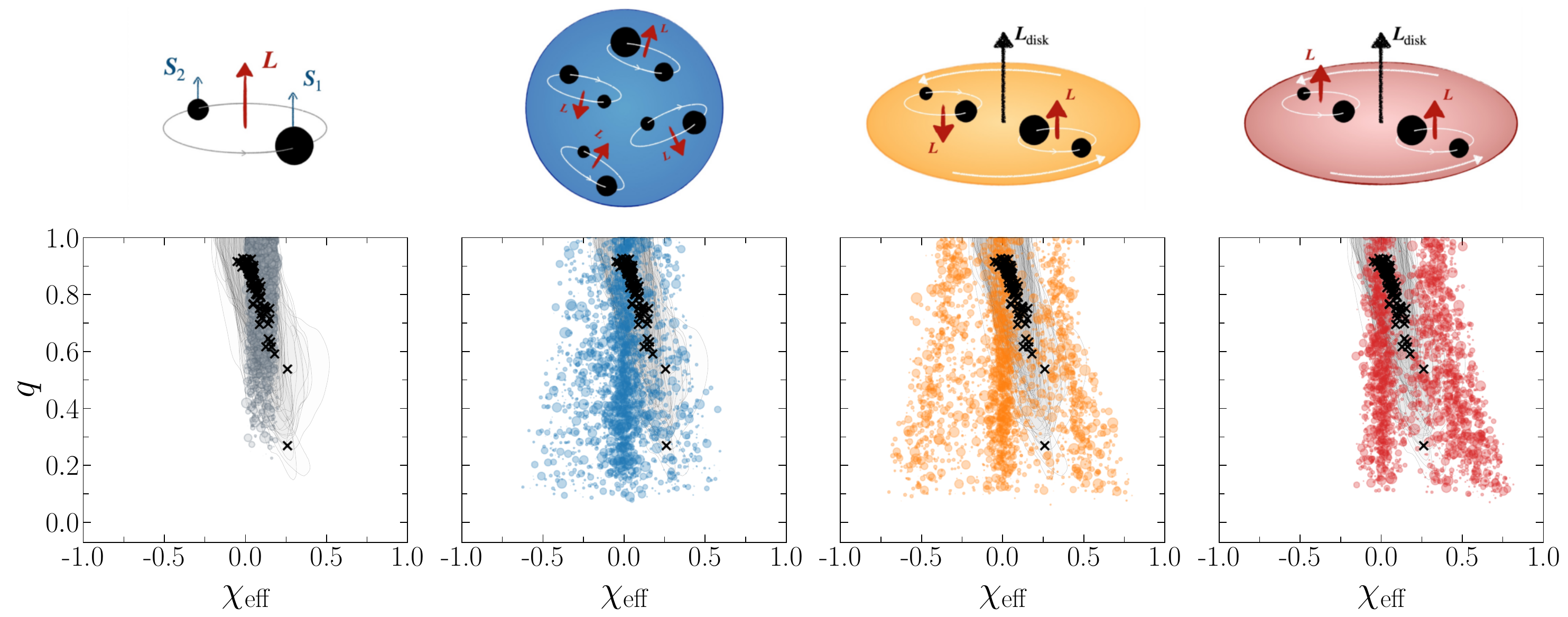}
		\caption{%
		Mass ratios $q$ and effective spins $\chi_{\rm eff}$ for binaries formed in the galactic fields (leftmost panel in gray), or in dense environments with three different symmetries. From left to right, we consider hosts with spherical symmetry (blue), cylindrical symmetry with both prograde and retrograde orbits (orange), and cylindrical with only prograde orbits (red). The parameters of the model described in Sec.~\ref{sec:model} are set to $\alpha=-3.5$, $\beta=1.1$,  $\gamma=-2.3$, $\chi_{\rm max}=0.2$, $f_{\rm disk}=0.2$, and $\lambda=1$. %
		The size of the markers is linearly proportional to the LIGO detection probability $p_{\rm det}$. Black crosses and gray contours indicate the one-dimensional medians and $90\%$ credible intervals of the GW-event posteriors reweighted to a population-informed prior that allows for the correlation~\cite{2023PhRvX..13a1048A}. The two-component model explored in this paper is a combination of the leftmost and rightmost populations of this figure. %
		}
		\label{fig:multiple_scatter}
	\end{figure*}

In this paper, we point out that the \emph{symmetry} of the astrophysical environment where BH mergers take place could play a pivotal role in explaining the observed $q-\chi_{\rm eff}$ correlation.  The key idea behind our study is illustrated in Fig.~\ref{fig:multiple_scatter}, where we contrast toy populations with and without hierarchical mergers and we consider different symmetries of the environment. 

Stellar clusters are, at least approximately, spherically symmetric. In the absence of a preferred direction, the BH spin orientations are expected to be distributed isotropically, which implies that positive and negative values of $\chi_{\rm eff}$ are equally probable ---a feature that can actually be used to put a limit on the fraction of hierarchical mergers~\cite{2020PhRvD.102d3002B,2022ApJ...935L..26F}.
For this reason, BH mergers in spherically symmetric environments cover a wedge in the $q-\chi_{\rm eff}$ parameter space (blue distribution in Fig.~\ref{fig:multiple_scatter}), with unequal masses paired to a wide range of effective spins covering both positive and negative values. %

Besides clusters, another promising environment to host hierarchical mergers are accretion disks surrounding active galactic nuclei (AGN). These systems are approximately axisymmetric, with a preferential direction set by the orbital angular momentum of the disk. At the toy-model level, one can naively assume that the
orbital angular momenta of stellar-mass BH binaries embedded in the disk will either (i)  coalign and counteralign, or (ii) strictly coalign with the symmetry axis~\cite{2005MNRAS.363...49K,2020MNRAS.494.1203M}. %
In the first case (yellow distribution in Fig.~\ref{fig:multiple_scatter}), hierarchical mergers have the same qualitative features highlighted for clusters, allowing for both positive and negative values of $\chi_{\rm eff}$, but with characteristic ``gaps'' between subpopulations of different generations. Instead, assuming coalignment with the external angular momentum of the disk (red distribution in Fig.~\ref{fig:multiple_scatter}) suppresses the left ``wing'' of the population and produces a negative correlation between $q$ and $\chi_{\rm eff}$. The gap between the central and the right subpopulations in the red distribution can be filled by binaries formed in isolation, which are expected to have mostly positive values of $\chi_{\rm eff}$ (gray distribution in Fig.~\ref{fig:multiple_scatter}).

Interestingly, AGN disks are the playground for what is perhaps the only astrophysical study to date looking for a possible origin of the $q-\chi_{\rm eff}$ correlation~\cite{2022MNRAS.514.3886M}. There %
the authors proposed numerous phenomenological, and admittedly tuned, considerations to suppress specific regions of the $q-\chi_{\rm eff}$ parameter space from their previous models~\cite{2020MNRAS.494.1203M}. %

The rest of this paper further explores the following questions:
\begin{itemize}
\item Can the symmetry of the environment
  explain, at least qualitatively, the observed $q-\chi_{\rm eff}$ correlation?
\item Looking ahead, could the mass-spin correlation of BH binaries be used to infer the symmetry of the astrophysical environments hosting BH mergers?
\end{itemize}
In Sec.~\ref{sec:model} we present a simple but concrete implementation of this idea. In Sec.~\ref{sec:results} we attempt a comparison with the LIGO-Virgo data. In Sec.~\ref{sec:outlook} we draw our conclusions and present possible directions for future research. %
  
\section{A simple model}
\label{sec:model}
We present a simplified set of prescriptions to explore the correlation between $q$ and $\chi_{\rm eff}$.
Our goal here is not to develop a complete model to fully explain current GW data or to provide Bayesian population fits.
Rather, we wish to explore some key physical ingredients that could produce a correlation at least qualitatively similar to what we currently observe.

\subsection{Building the populations}

The BHs observed by LIGO and Virgo
might be coming from {multiple formation channels with presumably comparable detection rates~\cite{2021ApJ...910..152Z}. It is therefore unlikely that the entire population of observable systems can participate in hierarchical mergers~\cite{2021ApJ...915L..35K,2022PhRvD.106j3013M}, which only occur in a subset of these channels~\cite{2021NatAs...5..749G}.} %
We thus consider a two-component model.  One can think of the first component as a proxy for isolated binaries formed in the galactic field, while the second component contains hierarchical mergers in an axisymmetric, disk-like setting. While we refer to our sources as ``field'' and ``disk,'' we stress that this is nothing more than a flexible setup to model their qualitative behavior.  The mixing fraction $f_{\rm disk} \in [0,\,1]$ quantifies the relative presence of mergers in the disk ($f_{\rm disk}$) %
and field ($1-f_{\rm disk}$) components. 
We assume a fiducial value $f_{\rm disk} = 0.2$ because {there is no strong evidence} that the majority of mergers originate from the AGN channel~\cite{2021ApJ...910..152Z, 2023arXiv230609415V},
but we have verified that our results are solid under variations of this parameter. %

All field BHs are of first generation (1g), while disk binaries can participate in hierarchical mergers of the $N$g$+1$g type with $N>1$~\cite{2019PhRvL.123r1101Y}. These are ``chain accretion'' episodes where an initial BH accretes $N$ objects from an available reservoir of 1g BHs.
In the context of AGN disks, the occurrence of such events is motivated by the potential presence of migration traps~\cite{2016ApJ...819L..17B,2020ApJ...898...25T}: locations in the disk where viscous drag pushes the inner perturber outward and the outer perturber inward. If/when a BH reaches a trap, it is expected to act as a catalyzer and accrete other objects that are brought to the same location by the disk dynamics~\cite{2019PhRvL.123r1101Y}. 
{Crucially, these migration-trap chains do not make up the totality of mergers in AGN disks. Additional $1$g$+1$g mergers are expected to take place in different regions of the disk~\cite{2020MNRAS.498.4088M}, and $N$g$+N$g mergers are also predicted to be present, although with a lower rate~\cite{2021ApJ...908..194T, 2020MNRAS.494.1203M}. This is an important caveat that should be taken into account when associating a physical meaning to $f_{\rm disk}$. } %

For the field component, we make the following assumptions. The primary masses $m_1 \in \left[5,\,50 \right]\, M_\odot$ have a distribution $p(m_1) \propto m_1^{\alpha}$,  the secondary masses have $p(m_2 | m_1)\propto m_2^\beta$ over the interval $m_2 \in \left[5\, M_\odot,\,m_1 \right]$, and the spin magnitudes are distributed uniformly in the range $\chi \in \left[0,\, \chi_{\rm max} \right]$. %
We consider fiducial values of $\alpha = -3.5$ and $\beta=1.1$ inspired by current GW observations~\cite{2023PhRvX..13a1048A}, and we vary $\chi_{\rm max}$  in our parameter-space exploration (cf. Sec.~\ref{sec:results}).
We assume the field BH spins to be perfectly aligned with the orbital angular momentum of the binary ($\theta_{1,2}=0$). This 
is a simplifying but reasonable assumption that neglects, among others, the effect of natal kicks~\cite{2000ApJ...541..319K,2016ApJ...832L...2R, 2018PhRvD..98h4036G}.  

For the disk component undergoing hierarchical mergers, we sample the 1g BH masses according to $p(m) \propto m^{\gamma}$, and the spin magnitudes  uniformly in the same intervals considered above. We explore two possible values of $\gamma$: $\gamma= -2.3$, motivated by the
Kroupa initial mass function~\cite{2001MNRAS.322..231K}, and $\gamma=-1$, %
motivated by studies showing that disk dynamics may harden the BH mass spectrum~\cite{2019ApJ...876..122Y}. %

The details of the $N$g+$1$g merger series formed in  migration traps %
depend on the host properties, including AGN lifetime, accretion efficiency, and disk viscosity. For our simple model, we assume that each BH seed accretes 1g objects up to a maximum generation $N=N_{\rm max}$. %
Inspired by Ref.~\cite{2019PhRvL.123r1101Y}, we sample $N_{\rm max}$ from a Poisson distribution with mean $\lambda$: in practice, we are  encapsulating the numerous properties of the host in a single parameter $\lambda$ which controls the relative importance of hierarchical mergers, and thus their impact on the $q-\chi_{\rm eff}$ correlation. 
{For simplicity, we also neglect the role of gas accretion on the evolution of BH masses and spins.}

Our disk-like environments are defined by a preferential direction $\boldsymbol{L}_{\rm disk}$, 
which models the global orbital angular momentum of the disk. For each $N$g+$1$g merger chain, we assume that the angle $\theta_{L}$ between the angular momentum of the merging binaries $\boldsymbol{L}$ and  that of the disk $\boldsymbol{L}_{\rm disk}$ is distributed uniformly in cosine and bounded from above, i.e. $\theta_{L} \leq  \theta_{\rm max}$.  This is a crucial parameter of our model, as this angle controls the degree of symmetry of the environment. An axisymmetric host with coaligned binaries (red distribution in Fig.~\ref{fig:multiple_scatter}) corresponds to $\theta_{\rm max} = 0$,  while a cluster-like environment with isotropic spin directions (blue distribution in Fig.~\ref{fig:multiple_scatter}) corresponds to tilt angles distributed uniformly in cosine up to $\theta_{\rm max}= \pi$.

As for the spin directions, we assume all 1g BHs from the disk component to be distributed isotropically, as these are presumably captured from the surrounding environment~\cite{2020ApJ...898...25T}. %
For the $N$g BHs, we use numerical-relativity fitting formulas to estimate the remnant mass~\cite{2012ApJ...758...63B} and spin magnitude~\cite{2016ApJ...825L..19H}, as implemented in Refs.~\cite{2016PhRvD..93l4066G, 2023PhRvD.108b4042G}. We assume that the remnant spin is parallel to the total angular momentum of the binary $\boldsymbol{J} = \boldsymbol{L} + \boldsymbol{S}_1 +  \boldsymbol{S}_2$ before merger~\cite{2016ApJ...825L..19H, 2009ApJ...704L..40B}, where $ \boldsymbol{S}_i=m_i^2\chi_i$ are the BH spins. This yields%
\begin{align}
\theta_{\rm f} = \arccos \left(\frac{L + S_1 \cos\theta_1 + S_2 \cos\theta_2}{J} \right)\,.
\label{thetaf}
\end{align}
At the next merger in the $N$g+$1$g series, the tilt angle of the $N$g BH is equal to the $\theta_{\rm f}$ angle from the previous merger. The azimuthal spin angles are resampled isotropically, as they are degenerate with the orbital phase. %
In Eq.~(\ref{thetaf}), we estimate the orbital angular momentum from the Newtonian expression $L=m_1 m_2 \sqrt{r/M}$ evaluated at a fiducial separation $r=10 M$ before plunge, where $M=m_1+m_2$ is the total mass. This roughly corresponds to the breakdown of the post-Newtonian approximation, within which angular momenta can be added without taking into account the full complexity of general relativity. We have verified that this specific choice does not impact our results, which remain largely unaffected even for values of the orbital separation as small as $r\simeq 3 M$. %

BH remnants could  be ejected from their astrophysical host by recoils imparted during the merger process (the so-called ``BH kicks''), which in turn prevent the occurrence of hierarchical mergers~\cite{2019PhRvD.100d1301G}. Kick ejection is unlikely to play a relevant  role in AGN disks. Typical orbital velocities at the locations of the migration traps are of $\mathcal{O}(10^4)$ km/s~\cite{2016ApJ...819L..17B} ---hardly perturbed by typical BH kicks, which are of $\mathcal{O}(100)$ km/s~\cite{2018PhRvD..97j4049G}.  ``Superkicks'' of $\mathcal{O}(1000)$ km/s are possible~\cite{2007PhRvL..98w1102C,2007PhRvL..98w1101G} but very rare, because they require highly fine-tuned binary configurations.  We simply assume that BH remnants do not leave their hosts.

\begin{figure*}
		\centering
		\includegraphics[width=2\columnwidth]{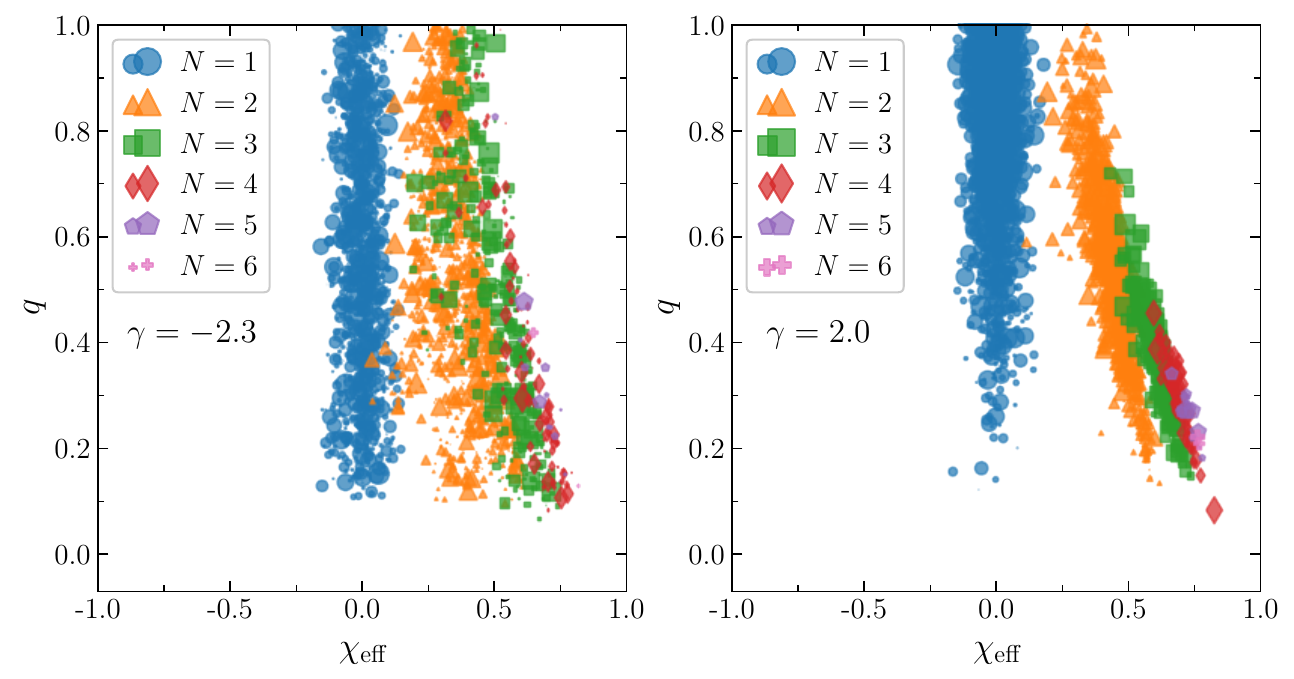}
		\caption{Disk-component of the population of BH binaries predicted by our model highlighting the contribution of the different merger generations. {We show the population predicted assuming $\alpha=-3.5$, $\beta=1.1$, $\chi_{\rm max}=0.2$, $f_{\rm disk}=0.2$, and $\lambda=1$. In the left panel we assume $\gamma=-2.3$, as in the rightmost panel of Fig.~\ref{fig:multiple_scatter}, while in the right panel we set $\gamma=2$.  }Colors and markers differentiate the $N$th BH generation in the occurring $N$g+$1$g merger chains. The size of the markers is linearly proportional to the LIGO detectability $p_{\rm det}$.  %
		} 
		\label{fig:high_generations}
	\end{figure*}

\begin{figure}
		\centering
		\includegraphics[width=\columnwidth]{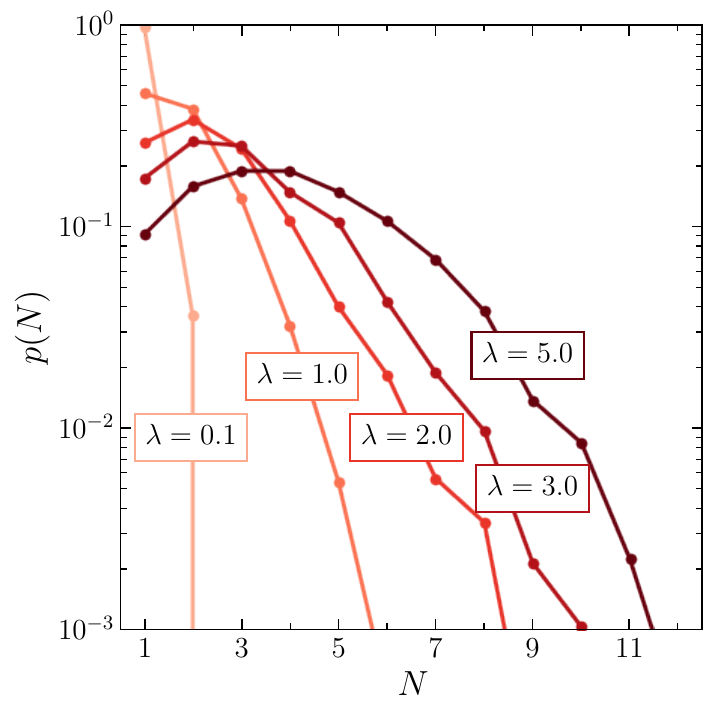}
		\caption{Normalized detectable fraction of events in each generation $p(N)$ for different choices of the Poisson parameter $\lambda$. Darker (lighter) colors correspond to larger (smaller) values of $\lambda$. %
		{We only show the disk sub-population of our two-component model and assume fiducial values $\alpha=-3.5$, $\beta=1.1$, $\gamma=-2.3$, $\chi_{\rm max}=0.2$ and $f_{\rm disk}=0.2$.}}
		\label{fig:high_generations2}
	\end{figure}

\subsection{Detectability}

Rather than generating populations with a fixed number of initial BHs, we keep assembling binaries until the cumulative
detection probability $p_{\rm det}$ ~\cite{1993PhRvD..47.2198F,2015ApJ...806..263D} reaches a pre-determined threshold $\sum_i{p_{{\rm det},i}} =1000$ (this specific number is not important and was set to obtain sufficiently large statistics when plotting results). This allows us to compare different sets of model parameters on equal footing.
We assign to each binary a redshift value $z$ extracted uniformly in comoving volume and source-frame time, namely %
$p(z) \propto (\dd V_c / \dd z) / (1 + z)$ assuming the \textsc{Planck18} cosmology~\cite{2020A&A...641A...6P}.
We consider a single interferometer with LIGO's \textsc{ZeroDetunedHighPower} noise curve~\cite{DCC_psd}, %
 simulate signals with the \textsc{IMRphenomD} waveform model~\cite{2016PhRvD..93d4007K}, and consider sources as detectable when their signal-to-noise-ratio is greater than $8$~\cite{2016ApJS..227...14A}. The detection probability $p_{\rm det}$ is estimated by marginalizing analytically over the extrinsic parameters~\cite{1993PhRvD..47.2198F} as implemented in the \textsc{gwdet} package~\cite{gwdet}. For computational efficiency, we neglect spin effects when computing $p_{\rm det}$, as these provide a subdominant contribution in the context of the highly simplified astrophysics of our model~\cite{2015ApJ...806..263D,2018PhRvD..98h4036G}. We have verified that this is a reasonable approximation by performing selected runs using the machine-learning classifier from Ref.~\cite{2020PhRvD.102j3020G}, which includes spin effects at the price of a higher computational cost. For the set of parameters adopted in Fig.~\ref{fig:multiple_scatter}, the difference in $p_{\rm det}$ is $\lesssim 0.14$ in $90\%$ of the cases, and our main results are essentially unchanged.

\section{Reproducing the observed correlation}
\label{sec:results}

We first analyze how different generations of binaries populate the $q-\chi_{\rm eff}$ plane. We then proceed to compare our results against the distributions predicted by current GW observations. Finally, we point out relevant caveats of our investigation, including the role of $\theta_{\rm max}$. Unless specified otherwise, %
we set $\theta_{\rm max}=0$. %

\subsection{Model predictions}

The left panel of Fig.~\ref{fig:high_generations}
 shows binaries in the disk component for a fiducial model  with $\alpha=-3.5$,  $\beta=1.1$, $\gamma=-2.3$,  $\chi_{\rm max}=0.2$, $f_{\rm disk}=0.2$, and $\lambda=1$. This is the same population shown in red in Fig.~\ref{fig:multiple_scatter}. {The right panel shows a model variation with $\gamma=2$ (cf. Sec.~\ref{sec:caveats}). In particular,} we highlight the BH generation $N$ of the $N$g+$1$g merger chains and show how these populate different regions in the $q-\chi_{\rm eff}$ plane.

Initial $1$g$+1$g binaries have $q\in \left[0.1, \, 1\right]$ and $\chi_{\rm eff}\in \left[-\chi_{\rm max}, \, \chi_{\rm max}\right]$, %
the latter being a direct consequence of Eq.~(\ref{definitions}) with isotropic spin directions.
As the seed BH undergoes subsequent mergers, binaries present, on average, larger values of $\chi_{\rm eff}$ and smaller values of $q$.  %
The lack of $N\geq 2$ binaries with $\chi_{\rm eff} < 0$ indicates that one merger is sufficient to align the spin of the newly formed $2$g BHs with the orbital angular momentum of the binary, and thus with the angular momentum of the disk.

\begin{figure*}
		\centering
		\includegraphics[width=2\columnwidth]{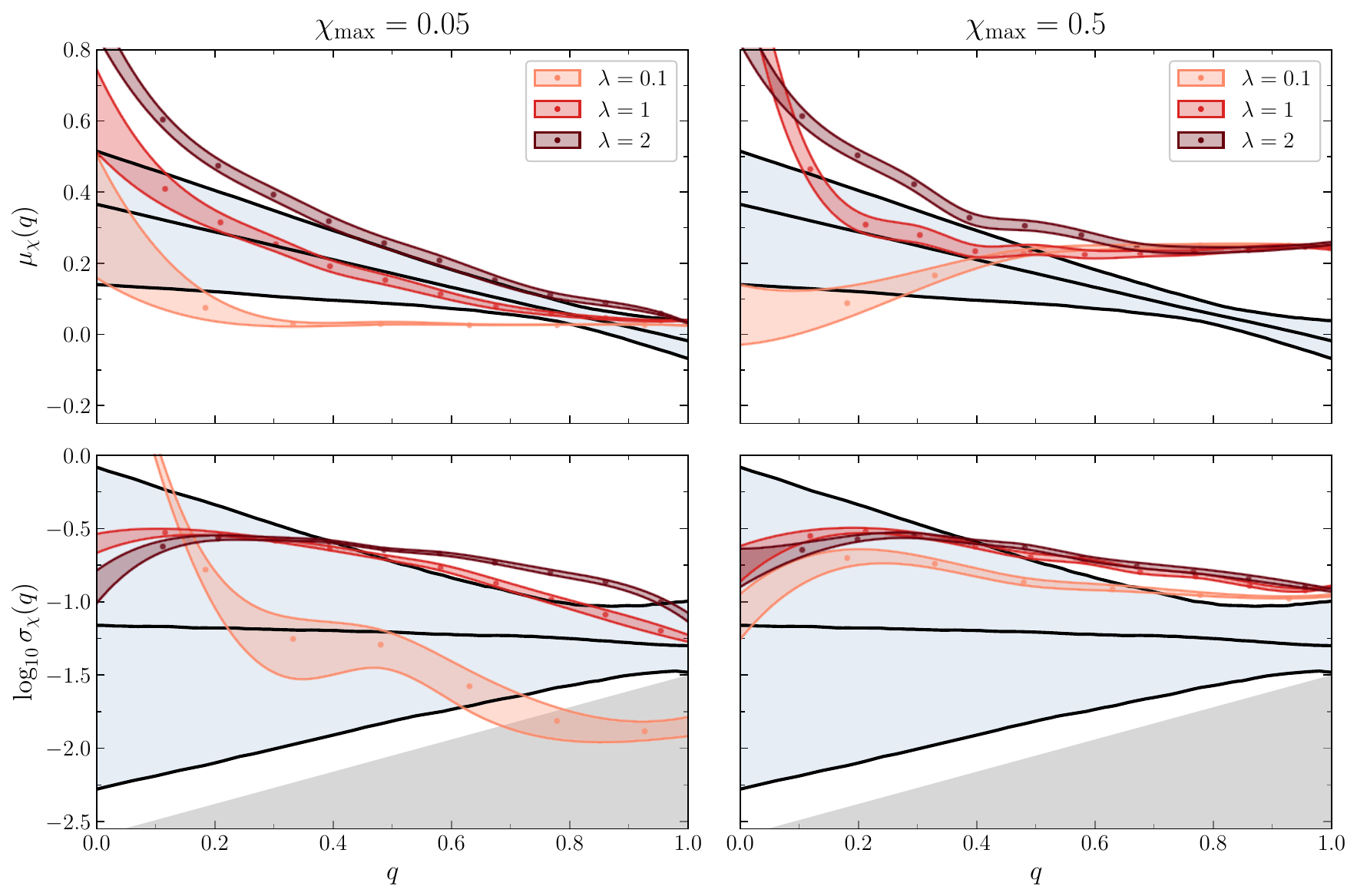}
		\caption{%
			Mean $\mu_\chi$ (top panels) and standard deviation $\sigma_{\chi}$ (bottom panels) of the $\chi_{\rm eff}$ distribution as a function of $q$. Black solid lines indicate medians and $90\%$ credible intervals from current GW data~\cite{2023PhRvX..13a1048A}. The gray region in the bottom panels is excluded by their choice of priors.
Our models are shown in shades of red, corresponding to means of the BH-generation distribution $\lambda=0.1,\, 1, \, 2$ (lighter to darker). The left (right) panels assume  
that the largest spin magnitude of first generation BHs is $\chi_{\rm max} = 0.05 \, (0.5)$. %
The remaining model parameters have been set to 
$\alpha=-3.5$, $\beta=1.1$, $\gamma=-2.3$, and $f_{\rm disk} = 0.2$. Colored circles indicate our point estimates in each bin. To quantify statistical uncertainties, the red shaded regions indicate $90\%$ confidence intervals on our predictions as estimated by bootstrapping and smoothed with a spline interpolation.%
		}
		\label{fig:comparison}
\end{figure*}

The relative rate of detectable sources steeply decreases with $N$. This is highlighted in 
{Fig.~\ref{fig:high_generations2},} where we show the cumulative detection probability of a given generation normalized to the total detection probability: %
\begin{equation}
  \label{eq:p}
  p(N) = \frac{ \sum_{i} p_{{\rm det},i} \; \mathcal{I}({\rm gen}_i= N) }{\sum_i p_{{\rm det},i} }\,,
\end{equation}
where $\mathcal{I}$ is an indicator function that is equal to $1$ if the generation of the sample $i$ is equal to $N$, and zero otherwise.
Larger values of $\lambda$ imply BHs of higher generations.
The observed trend is due to both our Poissonian assumption and to the sensitivity of ground-based detectors: higher-generation binaries present, on average, lower mass ratio, and are thus harder to detect.  %
The BH spectrum peaks at $N\simeq \lambda$, as expected.

Despite their reduced rate, our {fiducial} model predicts that some detectable binaries with $N\gtrsim 3$ should populate the region of the parameter space with $q \lesssim 0.4$ and $\chi_{\rm eff} \gtrsim 0.5$.
It also predicts an apparent excess of binaries with close-to-equal masses and moderate effective spins  (left panel of Fig.~\ref{fig:high_generations}).  %
The latter is, at least partially, a consequence of sampling all $1$g BH masses  from the same mass function with a uniform pairing probability (cf. Sec.~\ref{sec:caveats}). %
At present, GW data~\cite{2023PhRvX..13a1048A} do not provide significant support in either of these portions of the $q-\chi_{\rm eff}$ plane. On the one hand, this implies that a more sophisticated model is necessary; on the other hand, it also indicates that the next observing runs might provide constraints on the BH mass function in accretion disks, which is uncertain~\cite{2018ApJ...866...66M,2020ApJ...898...25T}. %

\subsection{Comparison with GW data}
\label{sec:comparison}
	
Despite its extreme simplicity, our  model can reproduce the joint $q-\chi_{\rm eff}$ distribution observed in current GW data. This is shown in Fig.~\ref{fig:comparison}, where we compare its predictions against results from Ref.~\cite{2023PhRvX..13a1048A}. In that study, GW data were analyzed assuming a population prior where the effective spin is normally distributed with mean and variance that are linear functions of the  mass ratio~\cite{2021ApJ...922L...5C}:%
	\begin{align}
		p(\chi_{\rm eff} | q) &= \frac{1}{\sqrt{2\pi\sigma_\chi ^2(q)}} \exp\left\{-\frac{[\chi_{\rm eff} - \mu_\chi (q)]^2}{2\sigma_\chi ^2(q)} \right\}\,, \label{pcheffiq}\\
		\mu_\chi (q) &= \mu_0 - \mu_1 (1-q) \label{linearmu}\,,  \\
		\log_{10}\sigma_\chi (q) &= \log_{10}\sigma_0 - \log_{10}\sigma_1 (1 - q)\,. \label{linearsigma}
	\end{align}

We use samples of the population parameters   $\left\{ \mu_0, \, \mu_1, \sigma_0, \, \sigma_1 \right\}$ publicly released with Ref.~\cite{2023PhRvX..13a1048A} and look for a combination of our model parameters that is able to capture at least the overall trend (see Sec.~\ref{sec:caveats} for important caveats on this procedure). %
In particular, in Fig.~\ref{fig:comparison} we explore values $\lambda = 0.1,\, 1, \, 2$ and $\chi_{\rm max}=0.05, \, 0.5$ %
while setting $\alpha=-3.5$, $\beta=1.1$, $\gamma=-2.3$, and $f_{\rm disk} = 0.2$. We have verified that setting $\gamma=-1$, as motivated in Sec.~\ref{sec:model}, results in distributions that are largely indistinguishable. %

We divide our simulated sources in equispaced bins along the $q$ directions and compute the mean and standard deviation of $\chi_{\rm eff}$ for each bin; these are compared against the measured values of $\mu_{\chi}$ and $\sigma_\chi$, respectively. For each simulation, the number of bins is selected such that each bin contains at least $50$ entries. %
Errors %
on the bin counts are estimated by bootstrapping~\cite{2019sdmm.book.....I}. Crucially, the analysis of Ref.~\cite{2023PhRvX..13a1048A} reports the \emph{observable} population of BH binaries, not the \emph{observed} one. For an apple-to-apple comparison, we exclude from our populations binaries with $p_{\rm det}=0$ (because they are not observable), but otherwise include all sources with equal weight. In other words, binaries are not filtered by detectability as long as $p_{\rm det}>0$. This is because selection effects have already been included by the authors of Ref.~\cite{2023PhRvX..13a1048A} and should not be double counted.

We find that the case with $\chi_{\rm max}=0.05$ %
and $\lambda=1$ is in reasonable agreement with the data. It is largely compatible with the measured values of both $\mu_\chi(q)$ and  $\sigma_\chi(q)$ when considering their Bayesian uncertainties. Larger (smaller) values of $\lambda$ tend to overestimate (underestimate) both $\mu_\chi(q)$ and $\sigma_\chi(q)$, while all cases with $\chi_{\rm max} = 0.5$ produce a large mismatch with the observations for $q \gtrsim 0.6$. 

In our populations, the predicted value of $\mu_\chi$ approaches $\chi_{\rm max}/2$ for $q\to 1$. For equal-mass systems, our population is mostly dominated by the isolated-binary component because $f_{\rm disk}< 1$ and $\beta>0$. From Eq.~(\ref{definitions}) with $q= 1$ and $\theta_{1,2}=0$ (which is assumed for our field binaries), one has $\chi_{\rm eff} = (\chi_1+\chi_2)/ 2$. Both spin magnitudes $\chi_{1,2}$ are distributed uniformly in $[0,\chi_{\rm max}]$, which implies that the expectation value of $\chi_{\rm eff}$ is equal to $\chi_{\rm max}/2$. Although this limit is not exactly reproduced in our populations because of a subdominant fraction of disk binaries with comparable masses and spins that are not necessarily aligned, we predict that the value of the effective spin for equal-mass binaries might be a relatively clean observable related to the maximum BH spin formed during stellar collapse~\cite{2019ApJ...881L...1F,2020A&A...636A.104B}.

\subsection{Caveats}
\label{sec:caveats}
While suggestive of a connection between the observed $q-\chi_{\rm eff}$ correlation and the symmetry of the astrophysical environment in which mergers take place, our exploration has some important caveats. 

First, 
we are not performing a rigorous statistical fit to identify the set of model parameters that best matches the data. 
While hierarchical Bayesian analyses~\cite{2019MNRAS.486.1086M, 2022hgwa.bookE..45V} are now standard practice in the field, our model is admittedly too simple, to the point that using such a detailed methodology would obscure the key trends. That is, the fit would most likely converge somewhere, but stretching its interpretation (as is sometimes done in the literature) would not, in our opinion, be appropriate. Instead, we opted for a simpler comparison which is in line with the simplicity of the model. This is sufficient for the main goal of this study, namely to point out that the symmetry of dense environments might represent an important ingredient to explain the observed correlation.
This intuition must be confirmed using both more accurate statistical approaches and more realistic astrophysical models.
 
For the same reason, we have restricted our exploration to the two-dimensional marginalized distribution of $q$ and $\chi_{\rm eff}$. The BH binary parameter space is of course higher dimensional, and a full comparison against the data should take additional features (total mass, redshift, other spin components) into account. Absolute rates, possibly in conjunction with other astrophysical probes such as AGN observations, could provide additional constraining power.   %

On the astrophysical side, our model contains only two components (here dubbed ``field'' and ``disk''), which is unlikely to be realistic~\cite{2021ApJ...910..152Z}. Within the disk component, assuming $N$g+$1$g merger chains relies on the presence of migration traps in AGNs, which is a topic of debate~\cite{2020ApJ...898...25T,2023arXiv230707546G}. As it is often the case, the inverse problem is more interesting (``can the $q-\chi_{\rm eff}$ correlation be taken as an indication of the existence of migration traps in AGN disks?''), though it requires more detailed modeling.

Requiring that the angular momentum of all binaries is strictly coaligned to that of the disk ($\theta_{\rm max}=0$) is arguably our strongest modeling assumption {(see e.g. Ref.~\cite{2021ApJ...923L..23W}).}
To this end, we briefly investigate how different values of $\theta_{\rm max}$ impact our results, thus exploring different degrees of symmetry of the environment.

\begin{figure}
  \centering
  \includegraphics[width=\columnwidth]{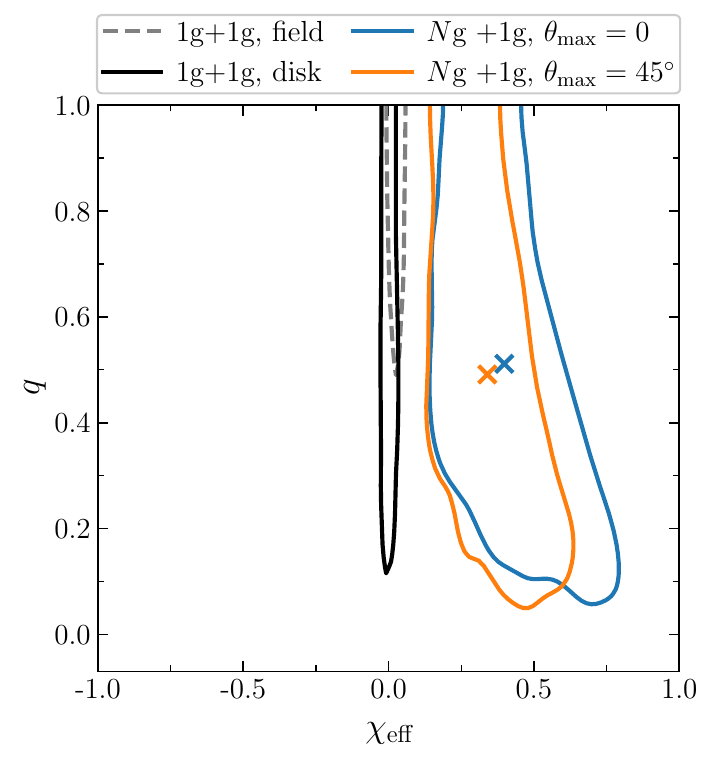}
  \caption{%
    Model predictions for different degrees of axisymmetry, as encoded in the parameter $\theta_{\rm max}$. We consider a model with parameters {$\alpha=-3.5$, $\beta=1.1$, $\gamma=-2.3$, $\chi_{\rm max} = 0.05$, $f_{\rm disk}=0.2$, and $\lambda=1$}. We show the $90\%$ contours of the resulting binaries,  weighted by the LIGO detectability $p_{\rm det}$. The dashed gray curve refers to the field component and the solid black line refers to $1$g+$1$g BHs (both of these are independent of $\theta_{\rm max}$). The two colored contours refers to detectable hierarchical mergers ($N>1$) assuming either $\theta_{\rm max}= 0$ as in the rest of the paper (blue) or $\theta_{\rm max}= 45^{\circ}$ (orange). 
            Crosses refer to the one-dimensional medians     of the respective distributions.
  }
  \label{fig:caveats}
\end{figure}

Figure~\ref{fig:caveats} compares the detectable population of disk binaries of different generations and different symmetries.
The distribution of $1$g$+1$g binaries %
is independent of $\theta_{\rm max}$.
On the other hand, hierarchical $N$g$+1$g binaries with $N>1$ are affected by $\theta_{\rm max}$, with larger values of $\theta_{\rm max}$ producing sources with smaller $\chi_{\rm eff}$ for a given $q$.
In our $N$g$+1$g chains, one BH merger is enough to align the spin of the remnant BH to $\boldsymbol{L}_{\rm disk}$. Therefore, if $\theta_{\rm max}=0$,  already at the second step of the sequence the angle between the angular momentum of the new binary and the spin of the remnant is $\theta_{1} \ssim 0$, which implies $\cos{\theta_{1}} \ssim 1$ (if the remnant is the primary BH in the new binary, which is the most likely case). Setting a non-zero value of $\theta_{\rm max}$ %
instead translates into values of $\cos{\theta_1} < 1$ and thus smaller values of $\chi_{\rm eff}$, regardless of the generation $N$.

Crucially, Fig.~\ref{fig:caveats} shows that our qualitative conclusions do not depend on the specific values of $\theta_{\rm max}$, as even increasing $\theta_{\rm max}$ from $0$ to $45^{\circ}$ causes a shift in the one-dimensional medians weighted by $p_{\rm det}$ as small as %
 $10\%$ (which is a subdominant variation when compared to the extent of the event posteriors from current data, cf. Fig.~\ref{fig:multiple_scatter}). 
We also verified that introducing such a degree of misalignment does not spoil the agreement with Ref.~\cite{2023PhRvX..13a1048A} highlighted in Sec.~\ref{sec:comparison}.

In conclusion, while \emph{strict} coalignment with $\boldsymbol{L}_{\rm disk}$ is not a crucial requirement of our model, considering hierarchical mergers with some \emph{preferential} coalignment is important, as allowing for counteralignment inevitably overpopulates the negative-$\chi_{\rm eff}$ region of the parameter space (Fig.~\ref{fig:multiple_scatter}), in tension with current observations. %

Finally, our treatment neglects mass segregation %
or, equivalently, a nontrivial BH pairing probability inside the accretion disk. In reality, more massive BHs are expected to migrate faster toward the putative migration traps~\cite{2019ApJ...878...85S}.

Within our model, we can mimic this effect by changing the mass spectral index of the disk component $\gamma$. A larger, positive value of $\gamma$ implies a top-heavy mass function that prefers more massive BHs. %
For instance, setting $\gamma =2 $  instead of $\gamma=-2.3$ heavily suppresses the presence of binaries with $q \ssim 1$ and $\chi_{\rm eff}\ssim 0.3 $, {see Fig.~\ref{fig:high_generations}.}
In particular, we do not find detectable $N$g$+1$g binaries with $N > 2$ for $q \gtrsim 0.7$, and just a handful of $2$g$+1$g events with $q \gtrsim 0.8$. %
While this goes in the direction of suppressing the top of the right wing in the red distribution of Fig.~\ref{fig:multiple_scatter}, which is indeed sparsely populated by current events, increasing $\gamma$ tends to overestimate both $\mu_{\chi}(q)$ and $\sigma_{\chi}(q)$.

More work and more physical models are needed to further investigate if and how mass segregation in AGN disks impacts the observed $q-\chi_{\rm eff}$ correlation, and thus (potentially) to constrain its occurrence with future data. %

\section{Conclusions}
\label{sec:outlook}

In this paper, we have constructed a toy model capable of reproducing the observed anticorrelation between the mass ratio $q$ and the effective spin $\chi_{\rm eff}$ of merging BHs. While surprising, this observational result withstood a large number of tests~\cite{2021ApJ...922L...5C,2022MNRAS.517.3928A,2023arXiv230715278A} and appears statistically solid. Additional points of scrutiny that should be better explored  %
include potential artifacts imposed by the underlying linear model of Eqs.~(\ref{linearmu}) and (\ref{linearsigma}), and subtle waveform systematics which might transfer biases from single-event analyses to population fits. %
That said, if the observed correlation is indeed of astrophysical nature, it offers a precious opportunity to constrain the pairing processes of merging BHs as well as their host environment. 

Our model is made of a bulk component of isolated field binaries and a smaller contribution of hierarchical $N$g$+1$g binaries assembled in axisymmetric, disk-like environments. Crucially, we assume that the orbital angular momenta of such a subpopulation share a preferential direction, thus imposing axisymmetry instead of the spherical symmetry most often assumed in the literature~\cite{2021NatAs...5..749G}.
Introducing this component appears to reproduce, at least qualitatively, the observed $q-\chi_{\rm eff}$ correlation without excessive fine-tuning. The required $N$g$+1$g merger chains are motivated by the occurrence of migration traps in AGN disks around supermassive BHs~\cite{2016ApJ...819L..17B}. This connection hints at the exciting possibility of constraining the fine details of accretion physics using GW data.

Despite some important caveats (Sec.~\ref{sec:caveats}), our model reproduces the observational trend  (i) without artificially boosting the magnitude of BH spins at core collapse~\cite{2019ApJ...881L...1F} ($\chi_{\rm max}=0.05$ in Sec.~\ref{sec:comparison}), (ii) without requiring that AGN disks constitute the dominant BH-binary formation channel ($f_{\rm disk}=0.2$ in Sec.~\ref{sec:comparison}), (iii) without assuming long hierarchical merger chains ($\lambda=1$ in Sec.~\ref{sec:comparison}) which would contradict observations~\cite{2021ApJ...915L..35K,2022PhRvD.106j3013M}, and (iv) without strict assumptions on the alignment process (i.e. on $\theta_{\rm max}$, see Sec.~\ref{sec:caveats}). While the details must be ironed out with full population fits and more realistic astrophysical setups, the generic trend highlighted in this paper appears to be solid with respect to our model variations, as long as a component of hierarchical mergers in axisymmetric environments is present. This is, in our opinion, the key, qualitative ingredient that might shed light on the astrophysical origin of the observed $q-\chi_{\rm eff}$ correlation.

\acknowledgments

We thank Daria Gangardt, Konstantinos Kritos, Luca Reali, Barry McKernan, and Saavik Ford for discussions.
A.S. and D.G. are supported by ERC Starting Grant No.~945155--GWmining, 
Cariplo Foundation Grant No.~2021-0555, MUR PRIN Grant No.~2022-Z9X4XS, 
and the ICSC National Research Centre funded by NextGenerationEU.
A.S. is supported by an Erasmus+ scholarship. 
D.G. is supported by Leverhulme Trust Grant No.~RPG-2019-350 and MSCA Fellowship No.~101064542--StochRewind.
E.B. and R.C. are supported by NSF Grants No. AST-2006538, PHY-2207502, PHY-090003 and PHY-20043, and by NASA Grants No.~20-LPS20-0011 and 21-ATP21-0010. 
E.B. is supported by the Italy-USA Science and Technology Cooperation program, supported by the Ministry of Foreign Affairs of Italy (MAECI).
Computational work was performed at CINECA with allocations through INFN, Bicocca, and ISCRA project HP10BEQ9JB, at the Advanced Research Computing at Hopkins (ARCH) core facility %
 supported by the NSF Grant No. OAC-1920103, and at the Texas Advanced Computing Center (TACC) at the University of Texas at Austin.

\bibliography{qchicorrelation}

\end{document}